\documentstyle[12pt]{article}

\catcode`@=11
\def\@cite#1#2{$^{\scriptscriptstyle
\mbox{\rm\scriptsize#1\if@tempswa , #2\fi}}$}

\def\mite{\@ifnextchar [{\@tempswatrue\@mitex}{\@tempswafalse\@mitex[]}}

\def\@mitex[#1]#2{\if@filesw\immediate\write\@auxout{\string\citation{#2}}\fi
  \def\@mitea{}\@mite{\@for\@miteb:=#2\do
    {\@mitea\def\@mitea{,}\@ifundefined
       {b@\@miteb}{{\bf ?}\@warning
       {Mitation `\@miteb' on page \thepage \space undefined}}%
\hbox{\csname b@\@miteb\endcsname}}}{#1}}

\def\@mite#1#2{$\mbox{\rm#1\if@tempswa , #2\fi}$}
\catcode`@=12

\def\onward{\addtocounter{section}{1} \setcounter{equation}{0} }

\newcommand{\eq}{\begin{equation}}
\newcommand{\en}{\end{equation}}
\newcommand{\ie}{{\it i.e.}}
\def\kite#1#2{$^{\mbox{\footnotesize \rm   #1-#2 \normalsize}}$}

\def\ZZ{Z\!\!\!Z}

\def\CC{I\!\!\!\!C}

\def\col{k}

\def\row{l}

\def\lam{\lambda}

\def\tchi{J}
\def\hchi{{\raise 0.8mm \hbox{$\chi$}}}
\def\bbox{\hat{c}}

\def\cI{{\cal I}}

\def\cR{{\cal R}}

\def\half{{\scriptstyle \frac{1}{2}}}
\def\sst{\scriptscriptstyle}
\def\sct{\scriptstyle}

\def\ta{\widetilde{a}}
\def\tb{\widetilde{b}}
\def\tc{\widetilde{c}}

\def\vev#1{\langle #1 \rangle}

\def\char{{\rm char}}

\def\sun{{\rm SU}(N)}

\def\spn{{\rm Sp}(N)}

\def\sunk{{\rm SU}(N)_{K}}
\def\sukn{{\rm SU}(K)_{N}}

\def\spnk{{\rm Sp}(N)_{K}}

\def\spkn{{\rm Sp}(K)_{N}}

\def\smlef{\hbox{$\sct ($}}
\def\smrig{\hbox{$\sct )$}}

\def\bspnk{{\bf {\rm\bf Sp}(N)_{K}}}
\def\uone{{\rm u} {\raise 0.25mm\smlef} {\sct 1} {\raise 0.25mm\smrig}}

\begin{document}
\setlength{\unitlength}{0.25cm}
\thispagestyle{empty}

\hfill{{\sc BRX-TH}-327}
\vspace{1.7cm}

\begin{center}
{\LARGE  Topological Landau-Ginzburg Matter
 from $\spnk$ Fusion Rings\footnote{Supported in part by
               the US Dept. of Energy under contract no.
               DE-AC02-76-ER03230}}\\[1.5cm]

{\hspace{-0.50in} \hbox{{\large \sl Michelle Bourdeau, Eli J. Mlawer,
         Harold Riggs, and Howard J. Schnitzer}} }

\vspace{0.5cm}
{ \sl
\begin{tabular}{c}
   Department of Physics \\
   Brandeis University \\
   Waltham, MA 02254
\end{tabular}  }

\end{center}

\vfill
\begin{center}
{\sc Abstract}
\end{center}

\begin{quotation}
We find and analyze the Landau-Ginzburg potentials whose
critical points determine
chiral rings which are exactly the fusion rings of $\spnk$ WZW models.
The quasi-homogeneous part of the potential associated with
$\spnk$ is the same as the quasi-homogeneous part
of that associated with ${\rm SU}(N+1)_{K}$, showing that these
potentials are different perturbations of the same Grassmannian
potential.

Twisted $N=2$ topological Landau-Ginzburg theories are derived from
these superpotentials. The correlation
functions, which are just the $\spnk$
Verlinde dimensions, are expressed as fusion residues.

We note that the $\spnk$ and $\spkn$ topological Landau-Ginzburg
theories are identical, and that while the $\sunk$ and
$\sukn$ topological Landau-Ginzburg models are not, they are
simply related.

\end{quotation}
\vfill
October 1991  \hfill

\setcounter{page}{0}
\newpage
\setcounter{page}{1}

\noindent {\bf 1. Introduction~}
\setcounter{section}{1}
{}From the beginning\cite{bpz} the Landau-Ginzburg (LG)
approach has provided
a useful picture of rational conformal field theories.\cite{rgzam}
More recently the Landau-Ginzburg formulation of $N=2$ superconformal
theories has yielded detailed results since the equation of motion
for the superpotential ${\rm d}W = 0$ is not subject to
renormalizations.\cite{mart}  A subset of the fields in such LG theories form
a closed, non-singular operator algebra, the chiral ring.
Such fields
have the property that their correlation functions are topological.
The non-topological fields can be removed by twisting these
theories to obtain LG models which are entirely
topological.\cite{witten,egu}  It is remarkable that there are deformations of
the $N=2$ superconformal theories to
non-conformal (but still $N=2$ supersymmetric) theories which
retain a closed topological ring.\cite{vvd,vafa} They
can also be twisted to obtain
topological LG models. These twisted $N=2$ models are important since they
serve as topological matter that can be coupled to
topological gravity.\kite{4}{9}

A particularly interesting example, a set of LG potentials
whose chiral rings are isomorphic to the $\sunk$ WZW fusion rings, has
been presented in ref.~\mite{gepner}. These potentials, considered
as the superpotentials of $N=2$ supersymmetric theories, are obtained
by perturbing the Grassmannian superconformal field theories. It is
possible to twist these perturbed
theories to obtain topological LG theories.\cite{intri}

One might expect, due to the prominence of the simply-laced Dynkin
diagrams in singularity theory, for example, that only the simply-laced
groups such as $\sunk$ would be amenable to this approach in a
natural way.  In this paper we show that all the constructions
mentioned above
apply just as naturally to the non-simply laced case of $\spnk$.
(Two recent papers\cite{warner,cresc}
appeared during the preparation of this letter which
suggest that LG potentials exist for arbitrary WZW fusion rings.)

Here we give the first detailed description of a
fusion ring of a non-simply laced group, $\spnk$, as a LG ring.
As an immediate application of this result, we construct
the twisted $N=2$ topological LG model by using these $\spnk$
potentials as superpotentials.  We find that the theory
characterized by an $\spnk$ potential is a deformation of the
same Grassmannian that can be perturbed (in a different way) to obtain
the ${\rm SU}(N+1)_{K}$
based LG models.  A further result is that the twisted versions
of the $\spnk$ and
$\spkn$ based topological LG theories are identical (on surfaces of any
genus). We also find that although the $\sunk$ and $\sukn$
theories are only identical on genus zero, their correlation
functions on arbitrary genus surfaces are closely related.

\vspace{0.2cm}
\noindent {\bf 2. Landau-Ginzburg Potentials for the $\bspnk$ Fusion Ring~}
\onward
First, we exhibit the $\spnk$ fusion ring as the quotient
of an unrestricted polynomial ring by an ideal.  We then present a
Landau-Ginzburg potential and show that its chiral ring
is the fusion ring.

The primary fields of
$\spnk$ ($N={\rm rank}\{\spn\}$)
are naturally described by the Young tableaux with at most
$N$ rows and $K$ columns. Let $\row_{i}$ denote the $i^{\rm th}$
row length and $\col_{i}$ the $i^{\rm th}$ column length of a given
tableau. The row lengths are
related to Dynkin indices by $\row_{i}=\sum_{j=i}^{N} a_{i}$.
The representation of $\spn$ with fundamental
highest weight $\Lambda_{j}$ is described by a tableau with a
single column of length~$j$.
Let $\hchi_{1},\ldots,\hchi_{N}$ denote the characters of these fundamental
representations.

In order to obtain the fusion ring as a quotient of a free ring in these
fundamental variables by an ideal, first consider the free ring
generated by the infinite set $\{\hchi_{i} ~\!| ~i=0,\ldots,\infty\}$.
In this context define the character of
an arbitrary tableau $\lambda$
with first row length $\row_{1}=s$
by the Giambelli type formula\cite{fh}
\eq
   \char(\lam) =
    \det \left|   \begin{array}{cccc}
\hchi_{\col_{1}}  &  (\hchi_{\col_{1} + 1} + \hchi_{\col_{1} -1}) &
   \ldots & (\hchi_{\col_{1}-1+s} + \hchi_{\col_{1}-s+ 1}) \\
     \vdots   &   \vdots   &  \vdots &  \vdots  \\
    \hchi_{\col_{i}-i+1} &  (\hchi_{\col_{i} - i + 2} + \hchi_{\col_{i} -i}) &
   \ldots & (\hchi_{\col_{i} - i + s} + \hchi_{\col_{i} -i-s+2}) \\
     \vdots    &   \vdots  & \vdots  & \vdots \\
     \hchi_{\col_{s}-s+1} &  (\hchi_{\col_{s} - s + 2} + \hchi_{\col_{s} -s}) &
   \ldots & (\hchi_{\col_{s}} + \hchi_{\col_{s} -2s+2}) \\
   \end{array}   \right|
\label{basdet}
\en
with $\hchi_{0}=1$ and $\hchi_{j} = 0$ for $j<0$.
An alternate
expression for $\char(\lambda)$ can be given in terms of
the tableaux with just one row.\cite{fh}  Let $\tchi_{i}$ denote the
character of a tableau consisting of a single
row of length $i$. Then for an arbitrary tableau with first {\it column}
length $\col_{1}=p$
\eq
  \char(\lam)= \det \left| \begin{array}{cccc}
           \vdots   &     \vdots      &  \vdots  &  \vdots   \\
    \tchi_{\row_{i} -i+1} &  (\tchi_{\row_{i} - i + 2} + \tchi_{\row_{i} -i}) &
   \ldots & (\tchi_{\row_{i} - i + p} + \tchi_{\row_{i} -i-p+2}) \\
         \vdots   &     \vdots      &  \vdots  &  \vdots
   \end{array}   \right|
\label{transdet}
\en
Even though a tableau with $p>N$ does not correspond to
an $\spn$ representation, these formulas define a character for all
such tableaux in terms of the $\hchi_{j}$ or $\tchi_{j}$.

The $\spnk$ fusion algebra operation is defined by
\eq
      \phi_{a} \phi_{b} = \sum_{c} {N_{ab}}^{c} ~\phi_{c}
\label{fusdef}
\en
where  the non-negative integers ${N_{ab}}^{c}$ can be calculated
as follows.
The product of the characters of
arbitrary tableaux can be computed in any
given case from the Pieri type formula,\cite{cking}
\eq
     \char (\lam) \char (\mu)
=  \sum_{\nu}
\char (~\!(\lam/\nu) \cdot (\mu/\nu) ~\!)
\label{mixal}
\en
where $(\alpha/\beta)$ denotes the sum of all tableaux $\gamma$ such
that $\alpha \in \beta \otimes \gamma$, and the dot indicates
the Littlewood-Richardson product.  This result follows ultimately\cite{cking}
from the definition of the characters in terms of the $\hchi_{j}$
given in eq.~\ref{basdet}.
Tableaux not corresponding to $\spnk$
primaries (\ie, those with more than $N$ rows  or $K$ columns)
{\it can} appear in eqs.~\ref{basdet} and~\ref{mixal}.

In order for eq.~\ref{mixal} to reproduce the $\spn$ tensor
ring, the characters of arbitrary
tableaux defined by the determinants above
must satisfy, in addition, a set of identities
(the {\it rank} modification rules) which allow
elimination of all tableaux
with more than $N$ rows.\cite{cking} This elimination
means that the characters of
tableaux corresponding to actual
$\spn$ representations are functions of a finite number of
fundamental variables, specifically, $\hchi_{0},\ldots,\hchi_{N}$.

The characters $\hchi_{0}, \hchi_{1}, \ldots$ are then connected with
certain elementary symmetric functions $E_{j}$ by the following
equation,\cite{fh} which holds for {\it all} $j$:
\eq
     \hchi_{j} = E_{j} - E_{j-2}    ~~~,
\label{charsym}
\en
with $E_{j} = 0$ for $j < 0$.
The $E_{j}$ with $j \geq 0$ are defined in terms of
the auxiliary variables $q_{i}$ ($i=1,\ldots,N$)
via the generating function\cite{fh}
\eq
    \sum_{j=0}^{\infty} E_{j} ~t^{j} =
     \prod_{i=1}^{N} (1+ q_{i} t) (1 + q_{i}^{-1}t)   ~~.
\label{symdef}
\en
It follows that $E_{0}=1$,
$E_{2N-j} = E_{j}$ for $j=0,\ldots,2N$, and $E_{j} =0$ if $j> 2N$.

{}From this and eq.~\ref{charsym} the generating characters $\hchi_{j}$
clearly satisfy the relations
\eq
    \begin{array}{rcc}
    \hchi_{N+1}            &  =      & 0 \\
   \hchi_{N+2}  + \hchi_{N}   &  =   &  0  \\
           ~              &  \vdots & ~ \\
\hchi_{j} + \hchi_{2N+2-j}  &  =      & 0 \\
       ~                  &  \vdots & ~ \\
    \end{array}
\label{rankgen}
\en
These relations generate the classical (rank) modification rules
mentioned above.
As a result, the free ring of polynomials in the fundamental
characters $\hchi_{0}, \hchi_{1},\ldots,\hchi_{N}$ generates the
classical tensor ring.

Equations~\ref{basdet} and~\ref{charsym} also imply that
the character of the single row tableau of length $i$,
$J_{i}$, can be written as the $i^{\rm th}$ complete
symmetric function of the variables $q_{i}$ via
the generating function\cite{fh}
\eq
    \sum_{j=0}^{\infty} J_{j} ~t^{j} =
     \prod_{i=1}^{N} {1 \over (1-q_{i} t) (1- q_{i}^{-1}t)}  ~~.
\en

One can impose a further set of identities (the {\it fusion}
modification rules\cite{cummins})
that allow the elimination of all characters
of tableaux with more than $K$ columns
in favor of those corresponding to primary fields of $\spnk$.
The ring structure of eq.~\ref{fusdef} is thus determined
by considering the ring defined in eq.~\ref{mixal}  modulo
the ideal whose irreducible
elements are exactly given by the modification rules.
Equivalently, this is the ring of polynomials in
the basic variables $\hchi_{i}, \ldots,\hchi_{N}$ modulo
the (fusion) modification rules, also written in terms of these
basic variables:
\eq
        \cR =      \CC[\hchi_{1},\ldots,\hchi_{N}]/\cI_{f} ~~.
\label{Fring}
\en
The following relations
generate this quantum or fusion ideal $\cI_{f}$ since they imply
the fusion modification rules (the proof of which will appear
elsewhere)
\eq
    \begin{array}{rcc}
    \tchi_{K+1}  &  =  & 0 \\
  \tchi_{K+2} + \tchi_{K} & = & 0 \\
           ~              &  \vdots & ~ \\
     \tchi_{K+N} + \tchi_{K-N+2} & = & 0 \\
    \end{array}
\label{genvan}
\en

The remarkable fact which we describe in this paper is that
the ring $\cR$ can be obtained from a Landau-Ginzburg
potential $V$, \ie,
\eq
        \cR = \CC [ \hchi_{1}, \ldots, \hchi_{N} ]/{\rm d}V  ~~.
\en
To express the potential in a compact form we use
the connection between the fundamental characters
$\hchi_{j}$ and $\tchi_{j}$ and the symmetric functions
described above.

It is then natural to consider (in the light of ref.~\mite{gepner})
\eq
     V_{N+K+1} = {1 \over N+K+1}
\sum_{i=1}^{N} (q_{i}^{N+K+1} + q_{i}^{-(N+K+1)})
\label{pot}
\en
as candidates for potentials for the $\spnk$ fusion ring.
Although eq.~\ref{pot} involves inverse powers of the auxiliary
variables, it is a polynomial in the fundamental variables
$\hchi_{j}$. (This is clear from eq.~\ref{dervan} below.)

First we define a generating functional for these potentials
\eq
   V(t) =  \sum_{i=1}^{N} \log (1 + q_{i} t)(1 + q_{i}^{-1} t) =
  \sum_{m=1}^{\infty} (-1)^{m-1} V_{m} t^{m}
\en
so that
\eq
    V_{m} = {1 \over m} \sum_{i=1}^{N} (q_{i}^{m} + q_{i}^{-m})  ~~~.
\en
Then a calculation similar to that of ref.~\mite{gepner} shows that
\eq
     {\partial V_{m} \over \partial E_{i}} =
     \sum_{{\sst 0 \leq j \leq 2N}} (-1)^{j+1} J_{m-j}
    {\partial E_{j} \over \partial E_{i}}   ~~~.
\en
Since $E_{j} = E_{2N-j}$ (from eq.~\ref{symdef}), we find
\eq
     {\partial V_{m} \over \partial E_{i}} = \left\{
   \begin{array}{cc}
   (-1)^{i+1} (J_{m-i} + J_{m+i-2N} ) & {\rm for~} 1 \leq i \leq N-1 \\
                         ~             &       ~  \\
      (-1)^{i+1} J_{m-i}    & {\rm for~} i=N
   \end{array}
    \right.
\label{dervan}
\en
Since $\det (\partial \hchi_{j} / \partial E_{i}) = 1$, we observe,
upon setting $m=N+K+1$, that the $N$ critical point conditions
of $V_{N+K+1}$
obtained from eq.~\ref{dervan} ($\partial V_{m}/\partial E_{i} = 0$)
exactly coincide with the
generators of the ideal in eq.~\ref{genvan}.
This shows that the
fusion ring of $\spnk$ is, in a natural way, a Landau-Ginzburg
ring of the potential $V_{K+N+1}$ when written in terms
of the fundamental variables $\hchi_{1},\ldots,\hchi_{N}$.

As an example, consider the potentials $V_{K+3}$ (\ie, $N=2$)
for ${\rm Sp}(2)_{K}$.
Because the polynomial in $q$,
\eq
      \prod_{i=1}^{N} (q-q_{i})(q-q_{i}^{-1}) = 0  ~~,
\en
is satisfied by any $q_{i}^{\pm 1}$, we can find a recursion relation for
the potentials in terms of the fundamental variables
$x=\hchi_{1}$ and $y=\hchi_{2}$,
\eq
    (k+4)V_{k+4} - x (k+3) V_{k+3} + (y+1)(k+2) V_{k+2} - x (k+1) V_{k+1} +
\left\{  \begin{array}{c}
      ~4 ~~~~k=0  \\
   k V_{k} ~~k>0
       \end{array}
   \right\}   = 0
\en
where $V_{1}= x$, $V_{2}=\half x^2 - y - 1$, and
$V_{3}={\sct {1 \over 3}} x^3 - xy$ are the proper
initial conditions.
If this equation is scaled by $x\rightarrow \lambda x$
and $y \rightarrow \lambda^2 y$, we obtain a recursion relation
for the quasi-homogeneous part which is identical to the recursion
relation for  the quasi-homogeneous part of the ${\rm SU}(3)_{K}$
potential. (In general, the same is true for $\spnk$ and
${\rm SU}(N+1)_{K}$.)

It will also be useful to describe the critical point
conditions of $V_{N+K+1}$ in eq.~\ref{dervan}
in terms of the auxiliary variables $q_{i}$. From eq.~\ref{pot}
we find that these conditions are given by
\eq
      q_{j}^{2(N+K+1)} = 1          ~~~~{\rm for~} j= 1,\ldots,N ~~.
\label{eqmot}
\en
If $q_{i} \neq q_{j}^{\pm}$ for $i\neq j$ and $q_{i}^2 \neq 1$,
then the Jacobian of the variable
change $\hchi_{j} \rightarrow q_{j}$ is non-singular.

The $\spnk$ fusion ring has a discrete symmetry generated by
the operation $\rho$, which
is just the map of representations induced
by the $\ZZ_{2}$ automorphism of the extended Dynkin diagam
of $\spnk$.
In terms of tableaux,
$\rho(i)$ is the representation whose tableau is the {\it complement}
in the $N$ by $K$ rectangle of the tableau of $i$.\cite{dual4}
Then for integrable $\spnk$ representations $a$, $b$, and $c$
(from eq. 3.10 of ref.~\mite{dual4})
\eq
       {N_{ab}}^{c} = {N_{\rho(a) b}}^{\rho(c)}
\label{comfus}
\en
Let us denote by $\bbox$ the representation $\rho(0)$,
whose tableau is the complement of
the identity, \ie ~the rectangular tableau with $N$ rows and $K$ columns.
Eq.~\ref{comfus} then implies that
the fusion rule for the field $\phi_{\bbox}$ with any
primary field $i$ satisfies
\eq
     \phi_{\bbox} \cdot \phi_{i} = \phi_{\rho(i)}  ~~~.
\en
The presence of this
symmetry permits the construction of a nontrivial topological
LG model whose chiral ring is exactly the $\spnk$ fusion ring.

\vspace{0.2cm}
\noindent {\bf 3. Twisted $\bf N=2$
Topological Field Theories Based on $\bspnk$~}
\onward
The potential in eq.~\ref{pot} is not quasi-homogeneous,
which is expected,
since $\spnk$ is a rational conformal field theory.
Under the scaling $\hchi_{j} \rightarrow \lambda^{j} \hchi_{j}$,
one finds that the quasi-homogeneous part of eq.~\ref{pot} is generated by
\eq
     (V_{N+K+1})_{\sst {\rm quasi-hom}} =
    { 1 \over N+K+1} \sum_{i=1}^{N} q_{i}^{N+K+1}
\en
which is the potential  of the Grassmannian
\eq
     { U(N+K) \over U(N) \times U(K)}
\en
(as well as being
related\cite{gepner} to the Kazama-Suzuki
coset ${\rm SU}(N+1)_{K}/\sun \otimes U(1)$)
so that the Landau-Ginzburg potentials of the $\spnk$ and ${\rm SU}(N+1)_{K}$
fusion rings are different perturbations of the {\it same }
$N=2$ superconformal field theory.

These results will now be used to construct a topological
Landau-Ginzburg model by twisting the $N=2$ Landau-Ginzburg model
associated with $\spnk$.
The starting point is the theory characterized by the
superpotential $W(\Phi_{i})$ of an $N=2$ supersymmetric
LG theory, where the $\Phi_{i}$ denote chiral superfields.
We take the superpotentials to be the functions
in eq.~\ref{pot} expressed in terms
of the fundamental fields (labeled by single
column Young tableaux) which
are then replaced by the chiral superfields,
\ie ~$\hchi_{i} \rightarrow \Phi_{i}$.

A finite number of
states are topological and form a closed ring
\eq
     \cR = \CC[ \Phi_{1},\ldots,\Phi_{N} ]/{\rm d}W
\en
which is isomorphic to the $\spnk$ fusion ring in eq.~\ref{Fring}.
This result shows that the remarkable correspondence\cite{ceco} between
deformations of $N=2$ superconformal models defined by
Landau-Ginzburg potentials and rational conformal field theories
also extends to the case of {\it non-simply laced} groups in
a natural way.

The twisted version of these $N=2$ theories contains {\it only}
these chiral primary fields, which are the topological fields.
The genus $g$ correlation functions with an insertion of
an arbitrary function
$F$ of chiral superfields $\Phi_{i}$, with superpotential $W$, is\cite{vafa}
\eq
      \vev{F(\Phi_{i})}_{g} = \sum_{{\rm d}W=0} h^{g-1} F(\Phi_{i})
\en
where the handle operator\cite{witten} is\cite{vafa} (using the
normalization of ref.~\mite{intri})
\eq
     h= (-1)^{N(N-1)/2} \det (\partial_{i} \partial_{j} W) ~~.
\label{vafone}
\en
The sum is over the critical points at which ${\rm d}W=0$. Here,
there is one critical point for each primary field of $\spnk$.
At the $a^{{\rm th}}$ critical point of ${\rm d}W = 0$ one has
$ \Phi_{i}(a) = S_{ia}/S_{oa}$, which forms a diagonal
representation of the fusion rules.
We now compute the Hessian $H$ of $W$ at these critical points:
\eq
     H= \det \left( {\partial^{2} W   \over \partial \hchi_{j} \partial
\hchi_{k}}
              \right)_{crit} =
\det \left( {\partial^{2} W   \over \partial E_{j} \partial E_{k}}
              \right)_{{\rm d}W =0 } =
{1 \over \Delta^{2}}
\det \left( {\partial^{2} W \over \partial q_{j} \partial q_{k}}
             \right)_{{\rm d}W = 0}
\label{Hfirst}
\en
where (using eq.~\ref{symdef})
\eq
     \Delta =
          \prod_{i=1}^{N} q_{i}^{-1}
   ~\prod_{i=1}^{N} (q_{i} - q_{i}^{-1})
 ~\prod_{i < j} [(q_{i} + q_{i}^{-1}) - (q_{j} + q_{j}^{-1})]
\en
A direct calculation using eq.~\ref{eqmot} gives
\eq
   \det \left( {\partial^{2} W \over \partial q_{j} \partial q_{k}}
             \right)_{{\rm d}W = 0}
    = 2^{N} (N+K+1)^{N} \prod_{j=1}^{N} q_{j}^{N+K-1} ~~~.
\label{Hfin}
\en

The Hessian $H$ can be related to the modular transformation
matrices $S_{oa}$ as follows. From ref.~\mite{kmod} we
know that
\eq
     S_{ab}
  = (-)^{ N(N-1)/2 } \left( 2 \over K+N+1  \right)^{N/2}  \det {\rm Mat}(a,b)
\en
where
\eq
   {\rm Mat}_{ij}(a,b)  = \sin
           \left( \pi \theta_i (a) \theta_j (b) \over K+N+1 \right),
\en
for $i,j = 1,\ldots,N$, and
\eq
  \theta_i (a) = \row_i (a) - i + N + 1, ~~~~~i=1, \ldots, N
\en
where $\row_{i}(a)$ are the row lengths of the tableau $a$.

Then with
\eq
    q_{j}(a) = e^{i\pi \theta_{j}(a)/(N+K+1)}
\label{auxcrit}
\en
we find that (the $a$ dependence of the $q_{j}$ will be
suppressed in what follows)
\eq
    S_{0a} = { (-i)^N \over [2(N+K+1)]^{N/2} }
   \prod_{i=1}^{N} (q_{i} - q_{i}^{-1})
 ~~\prod_{i < j} [(q_{i} + q_{i}^{-1}) - (q_{j} + q_{j}^{-1})]
\label{Sone}
\en
{}From eq.~\ref{auxcrit} we also find
\eq
   \prod_{j=1}^{N} q_{j}^{N+K-1} =
    \prod_{j=1}^{N} (-1)^{\theta_{j}(a)} q_{j}^{-2} =
     (-1)^{N(N + 1)/2} (-1)^{r(a)} \prod_{j=1}^{N}  q_{j}^{-2}
\label{Stwo}
\en
where $r(a)$ is the number of boxes of the tableau $a$.
Combining eqs.~\ref{Hfirst}-\ref{Hfin}, and eqs.~\ref{Sone}-\ref{Stwo}
we find
\eq
\det \left( {\partial^{2} W   \over \partial \hchi_{j} \partial \hchi_{k}}
              \right)_{a {\rm -th~crit}} =
   { (-1)^{N(N-1)/2} (-1)^{r(a)}  \over (S_{0a})^{2} }
\label{bigres}
\en
Therefore using eq.~\ref{vafone} we find that
\eq
      (h)_{a-{\rm th}} = {(-1)^{r(a)} \over (S_{oa})^2}  ~~.
\label{htwo}
\en
The value of the field $\Phi_{b}$ at the $a^{\rm th}$ critical
point is given by the character at that point
\eq
      \Phi_{b}(a) = \char_{a}(b) = {S_{ba} \over S_{0a}}    ~~.
\en
Since (from eq. 2.10 of ref.~\mite{dual4})
\eq
    S_{a\rho(b)} = (-1)^{r(a)} S_{ab}
\en
and since $\bbox=\rho(0)$
\eq
   \Phi_{\bbox}(a) = \char_{a}(\bbox) = {S_{\rho(0) a} \over S_{oa}} =
          (-1)^{r(a)}      ~~~.
\en
Therefore the handle operator in eqs.~\ref{vafone} and~\ref{htwo}
satisfies
\eq
     (h)_{a-{\rm th}} =  (S_{0a})^{-2}   \Phi_{\bbox}(a)
\label{hquest}
\en
for all critical points $a$.
Due to the non-singularity of the matrix $S_{ab}$ this uniquely
specifies the operators and we have $h =  h_{0} \Phi_{\bbox} $,
where the untwisted handle operator satisfies
$h_{0}(a) = S_{0a}^{-2}$.

Since $\Phi_{\bbox}$ is the unique chiral primary field with maximal
charge,
\eq
          \vev{\Phi_{\bbox}}_{g=0} = 1  ~~~.
\en
The topological metric for the theory based on the
$\spnk$ superpotential is
\eq
      \eta_{ij} = \vev{\Phi_{i} \Phi_{j}}_{g=0} = {N_{ij}}^{\bbox}
   = \delta_{j,\rho(i)}
\en
where $\rho(i)$ is the complement tableau of $i$ in the $N$ by
$K$ rectangle, while the topological metric in
the unperturbed theory (the Grassmannian) is
\eq
       \eta_{ij}^{(0)} = \delta_{j,{\rm comp}(i)}
\en
where ${\rm comp}(i)$ is again the complement tableaux of $i$ in
an $N$ by $K$ rectangle.
Thus the metric is unchanged upon deformation
\eq
     \eta_{ij} = \eta_{ij}^{(0)}
\en
as is required for a perturbed topological theory.\cite{vvd}

Note that in the twisted ${\rm SU}(N+1)_{K}$ theory, where
the $\ZZ_{N+1}$ discrete symmetry allows construction of a
nontrivial topological LG theory,\cite{intri} the topological
metric involves exactly the same operation of tableau complement.
The operation $\sigma$ that generates this symmetry adds a
row of length $K$ to the top of the tableau $a$,\cite{dual4}
so that it connects representations that are cominimally
equivalent (in the terminology of ref.~\mite{dual4}).
Then $\sigma^{N}(\overline{\imath})$ implements exactly the
tableau complement in an $N \times K$ rectangle.

\vspace{0.2cm}
\noindent {\bf 4. Rank-Level Duality\footnote{For a review and
references see ref.~\mite{dual}} of Topological Field Theories~}
\onward
The pair of
topological field theories built by twisting the $N=2$ models
associated with $\spnk$ and $\spkn$ Landau-Ginzburg potentials
are in fact equivalent.
For the (untwisted) $N=2$ supersymmetric (non-conformal) theories
characterized by these superpotentials associated with
rational conformal field theories, the correlation functions
of the chiral primary fields $\Phi_{i_{1}},\ldots,\Phi_{i_{n}}$
can be expressed in terms of the modular transformation matrices
\eq
    \vev{\Phi_{i_{1}},\ldots,\Phi_{i_{n}}}_{g}^{(u)} =
      \sum_{p}  {S_{op}}^{-2(g-1)}
        { S_{i_{1} p} \over S_{0p}} \ldots
      { S_{i_{n} p} \over S_{0p}}
\label{untwis}
\en
where the superscript $(u)$ indicates the untwisted theory.
The correlation functions in the twisted theory can be calculated
for any genus in terms of the untwisted theory by inserting $\Phi_{\bbox}$
\eq
      \vev{\Phi_{i_{1}} \ldots \Phi_{i_{n}}}^{(t)} =
    \vev{\Phi_{i_{1}} \ldots \Phi_{i_{n}} (\Phi_{\bbox})^{g-1}}^{(u)}
\label{twis}
\en

{}From  the identity\cite{dual4}
\eq
\begin{array}{rcl}
     ( S_{ab} )_{\spnk} & =   & ( S_{\ta ~\!\tb} )_{\spkn} ~~,
 \end{array}
\en
where $\tilde{a}$ denotes the transpose tableau of $a$,
and the one-to-one correspondence
of primary fields of $\spnk$ and $\spkn$ given by transposition,
it follows directly from eqs.~\ref{untwis} and~\ref{twis} that
\eq
   \vev{\Phi_{a_{1}} \Phi_{a_{2}} \ldots  \Phi_{a_{n}}}_{\spnk}
 =  \vev{\Phi_{\ta_{1}} \Phi_{\ta_{2}} \ldots  \Phi_{\ta_{n}}}_{\spkn}
\en
for correlation functions on any genus surface of the
twisted $\spnk$ theory.

To obtain the analogous results for $\sunk$ we can use\cite{nak}
(where $r(a)$ denotes the number of boxes in the tableau $a$)
\eq
       \begin{array}{rcl}
  (S_{ab})_{\sunk} &=&
  \sqrt{{K \over N}} e^{-2\pi i r(a) r(b)/NK}
           ~\!(S_{\ta ~\!\!\tb})_{\sukn} \\[3mm]
  S_{\sigma(a) b} & = & e^{-2\pi i r(b)/N }  S_{a b} ~~{\rm ~~~~in~}\sunk
   \end{array}
\en
in eq.~\ref{untwis}.
Lemma 1 of ref.~\mite{nak} allows one to relate
a sum over primaries in one theory to a sum over primaries in
the other at the cost of one factor of $N/K$ (which results
from the different sizes of the cominimal orbits in $\sunk$ and
$\sukn$) on the right hand side of eq.~\ref{sudual} below, so that
for untwisted $\sunk$ and $\sukn$ the duality of
the {\it non-zero} correlation functions of chiral primaries
on a genus $g$ surface is
\eq
    \vev{\Phi_{i_{1}} \ldots \Phi_{i_{n}}}_{{\sst \sunk}}
= \left( {N \over K} \right)^{g}
    \vev{\Phi_{\sigma^{p_{1}}(\widetilde{\imath}_{1})}
    \ldots \Phi_{\sigma^{p_{n}}(\widetilde{\imath}_{n})}}_{{\sst \sukn}}
\label{sudual}
\en
for {\it any} set of integers $p_{1},\ldots,p_{n}$ such that
$\sum_{i=1}^{n} p_{i} = - \Delta {\rm ~mod~} N$ with
$\Delta = \sum_{j=1}^{n} r(i_{j})/N$.  Here $\sigma^{p}(i)$ denotes
the $p-{\rm th}$ representation in the cominimal
orbit (cf. ref~\mite{dual4}). (That $\Delta$ is an integer follows
from properties of the $\sunk$ fusion ring.)
The fact that any set of integers
that sum to $-\Delta {\rm ~mod~} N$ can be chosen on the right hand
side of eq.~\ref{sudual} is an example
of the general symmetry of the Verlinde dimensions on
arbitrary genus surfaces under the discrete $\ZZ_{N}$
symmetry of $\sunk$.
For $g=0$ and $n=3$ we recover the special cases of
the duality of non-zero
fusion coefficients $N_{abc} = N_{\ta~\!\!\tb~\!\!\sigma^{N-\Delta}(\tc)}$
and their cominimal symmetry
$N_{\sigma^{p_{1}}(a) ~\!\!\sigma^{p_{2}}(b) ~\!\!\sigma^{p_{3}}(c)}=
N_{\sigma^{q_{1}}(a) ~\!\!\sigma^{q_{2}}(b) ~\!\!\sigma^{q_{3}}(c)}$
for any $p_{i}$ and $q_{i}$
such that $\sum p_{i} = \sum q_{i} ~{\rm mod~}N$, which
are described in refs.~\mite{nak},~\mite{dual2} and~\mite{dual4}.
If $g\neq 0$ then we do {\it not} have an exact equality due to
the different size of the cominimal equivalence
classes (alternately, of the orbits of the simple currents).
The correlation functions in the twisted $\sunk$ theory are
related to those of the twisted $\sukn$ theory by exactly the
same formula~\ref{sudual} except that the insertion
of $\Phi_{\bbox}$ in eq.~\ref{twis} shifts $\Delta$
to $\Delta + (g-1)K$ in the
condition  on the $p_{i}$ below eq.~\ref{sudual}.

Therefore the twisted $N=2$ Landau-Ginzburg models based on
$\spnk$ and $\spkn$ are exactly dual, and the twisted $N=2$
theories based on $\sunk$ and $\sukn$ are simply related, although
only exactly dual on genus zero.

\vspace{0.2cm}
\noindent {\bf 5. Concluding Remarks~}
\onward
In this paper we have explained how certain LG potentials have
chiral rings that exactly reproduce the $\spnk$ fusion ring.
As applications of this result, we also examined the twisted
$N=2$ topological LG theories given by interpreting these potentials
as superpotentials, and then demonstrated the presence of
rank-level duality in the $N=2$ topological LG theories
based on $\spnk$ and $\sunk$.

The extent to which the remarkable pattern of connections
between many disparate problems in low dimensional physics
in the $\sunk$ case\cite{ceco,spieg} is reproduced for $\spnk$ is
presently under investigation.

\noindent {\bf Acknowledgment~} H. J. Schnitzer thanks
M. Crescimanno, D. Gepner, and C. Vafa for useful conversations, and
E. Mlawer thanks H. Pendleton for helpful discussions.


\begin{thebibliography}{99.}
\bibitem{bpz}  A. A. Belavin, A. M. Polyakov, and A. B. Zamolodchikov,
              \sl Nucl. Phys. \bf B241 \rm (1984) 333.
\bibitem{rgzam} A. B. Zamolodchikov, \sl Sov. J. Nucl. Phys. \bf 44
                \rm (1986) 529.
\bibitem{mart} C. Vafa and N. P. Warner, \sl Phys. Lett. \bf 218B
               \rm (1989) 51; E. Martinec, \sl Phys. Lett. \bf 217B
               \rm (1989) 431; W. Lerche, C. Vafa, and N. P. Warner,
               \sl Nucl. Phys. \bf B324 \rm (1989) 427.
\bibitem{witten} E. Witten, \sl Commun. Math. Phys. \bf 117 \rm (1988) 353;
                  \bf 118 \rm (1988) 411; \sl Nucl. Phys. \bf B340 \rm (1990)
                  284.
\bibitem{egu}  E. Eguchi and S.-K. Yang, \sl Mod. Phys. Lett.
              \bf A5 \rm (1990) 1693.
\bibitem{vvd}   R. Dijkgraff, H. Verlinde, and E. Verlinde
                   \sl Nucl. Phys. \bf B352 \rm (1991) 59.
\bibitem{vafa} C. Vafa, \sl Mod. Phys. Lett. \bf A6 \rm (1991) 337.
\bibitem{kkli}   K. Li, \sl Nucl. Phys. \bf B354 \rm (1991) 711, 725.
\bibitem{dykg}   R. Dijkgraaf and E. Witten, \sl Nucl. Phys. \bf B342
               \rm (1990) 486.
\bibitem{gepner}  D. Gepner, ``Fusion rings and Geometry,''
                      ITP preprint NSF-ITP-90-184.
\bibitem{intri} K. Intriligator, ``Fusion Residues,''
     Harvard Preprint HUTP-91/A041, August 1991.
\bibitem{warner} D. Nemeschansky and N.P. Warner,
                 ``Topological Matter, Integrable Models and Fusion Rings,''
                 USC Preprint USC-91/031, October 1991.
\bibitem{cresc} M. Crescimanno, ``Fusion Potentials for $G_{k}$
                and Handle Squashing,'' MIT preprint CTP 2021, October 1991.
\bibitem{fh} W. Fulton and J. Harris, lecture notes (to
         be published by Springer-Verlag).
\bibitem{cking} R. King, \sl J. Math. Phys. \bf 12 \rm (1971) 1588.
\bibitem{cummins} C. Cummins, \sl J. Phys. A: Math. Gen. \bf 24 \rm (1991) 391.
\bibitem{dual4} E. J. Mlawer, S. G. Naculich, H. A. Riggs, and
                H. J. Schnitzer, \sl Nucl. Phys. \bf B \rm 352 (1991) 863.
\bibitem{ceco}  S. Cecotti and C. Vafa, ``Topological Anti-Topological
                 Fusion,'' Harvard preprint HUTP-91/A031 (1991);
                P. Fendley and K. Intriligator, in preparation.
\bibitem{kmod}  V. Kac and M. Wakimoto, \sl Proc. Nat. Acad. Sci.
                \bf 85 \rm (1988) 4956.
\bibitem{dual} S. Naculich, H. Riggs, H. Schnitzer, and E. Mlawer,
               ``A Quantum Generated Symmetry of Conformal and
               Topological Field Theory,'' Brandeis preprint BRX-TH-318,
               to appear in the proceedings of the NATO Advanced
               Research Workshop on ``Quantum Field Theory,
              Statistical Mechanics, Quantum Groups, and Topology,''
              University of Miami, January 1991.
\bibitem{dual2}  S. G. Naculich and H. J. Schnitzer,
                  \sl Phys. Lett. \bf B244 \rm (1990) 235;
                  \sl Nucl. Phys. \bf B347 \rm (1990) 687.
\bibitem{nak} A. Kuniba and T. Nakanishi,
            ``Level-rank duality in fusion RSOS models,''
              preprint, to appear in proceedings of \sl Int. Coll. on
              Modern Quantum Field Theory, \rm TIFR, Bombay, India.
\bibitem{spieg} M. Spiegelglas, ``Setting Fusion Rings in
                Topological Landau-Ginzburg,'' Technion preprint PH-8-91.
\end{thebibliography}
\end{document}